\documentclass[aps,prf,onecolumn]{revtex4-1}
\usepackage{amsmath}
\usepackage{graphicx}
\usepackage{color}
\usepackage{hyperref}
\usepackage{psfrag}
\usepackage{stmaryrd}
\usepackage{amssymb}
\usepackage{wasysym}
\usepackage{float}

\usepackage{geometry}
\begin{document}

\title{Two-dimensionalization of the flow driven by a slowly rotating impeller in a rapidly rotating fluid}

\author{Nathana\"el Machicoane}
\affiliation{Laboratoire FAST, CNRS, Universit\'e Paris-Sud,
Universit\'e Paris-Saclay, 91405 Orsay, France}
\author{Fr\'{e}d\'{e}ric Moisy}
\affiliation{Laboratoire FAST, CNRS, Universit\'e Paris-Sud,
Universit\'e Paris-Saclay, 91405 Orsay, France}
\author{Pierre-Philippe Cortet}
\affiliation{Laboratoire FAST, CNRS, Universit\'e Paris-Sud, Universit\'e
Paris-Saclay, 91405 Orsay, France}

\date{\today}

\begin{abstract}

We characterize the two-dimensionalization process in the
turbulent flow produced by an impeller rotating at a rate $\omega$
in a fluid rotating at a rate $\Omega$ around the same axis for
Rossby number $Ro=\omega/\Omega$ down to $10^{-2}$. The flow can
be described as the superposition of a large-scale vertically
invariant global rotation and small-scale shear layers detached
from the impeller blades. As $Ro$ decreases, the large-scale flow
is subjected to azimuthal modulations. In this regime,  the shear
layers can be described in terms of wakes of inertial waves
traveling with the blades, originating from the velocity
difference between the non-axisymmetric large-scale flow and the
blade rotation. The wakes are well defined and stable at low
Rossby number, but they become disordered at $Ro$ of order of 1.
This experiment provides insight into the route towards pure
two-dimensionalization induced by a background rotation for flows
driven by a non-axisymmetric rotating forcing.

\end{abstract}

\maketitle

\section{Introduction}

A key feature of rapidly rotating flows is the tendency towards
two-dimensionalization, a result known as the Taylor-Proudman
theorem~\cite{GreenspanBook,DavidsonBook2013,Godeferd2015}. For
asymptotically large rotation rates $\Omega$, this theorem states
that the fluid motion with characteristic time much larger than
the rotation period $\Omega^{-1}$, called {\it geostrophic flow},
is two-dimensional, invariant along the rotation axis (hereafter
called vertical by convention). For large Reynolds number and
moderate to small Rossby number (which compares the rotation
period $\Omega^{-1}$ to the turbulent turnover time), such slow,
large-scale, geostrophic flow may coexist with fast,
three-dimensional fluctuations in the form of inertial
waves~\cite{Leoni2014,Yarom2014,Campagne2015}. Inertial waves are
anisotropic, circularly polarized dispersive waves that propagate
because of the restoring nature of the Coriolis
force~\cite{GreenspanBook,McEwan1970,Machicoane2015}.

It was recently demonstrated by Gallet~\cite{Gallet2015} that
two-dimensionality can be reached exactly provided that the Rossby
number is smaller than a Reynolds-number-dependent critical value.
However, for most laboratory experiments, numerical simulations
and geophysical flows, the Rossby number is moderate, $Ro \simeq
10^{-2} - 1$, and the system is generally far from this asymptotic
two-dimensional state~\cite{Pedlosky1987,Vallis}. This departure
from two-dimensionality cannot be neglected in general: although
most of the kinetic energy is carried by the geostrophic flow,
three-dimensional fluctuations still remain responsible of most of
the energy dissipation~\cite{Campagne2015,Baqui2015}.

A question of practical interest is, for given initial or boundary
conditions, to what extent a turbulent flow in a rotating frame is
two-dimensional, and how the amount of two-dimensionality depends
on the Reynolds and Rossby numbers. This question has received
much attention, both
experimentally~\cite{Yarom2013,Campagne2014,Campagne2015} and
numerically~\cite{Bourouiba2012,Teitelbaum2012,Deusebio2014,Delache2014,Alexakis2015}.
Recently, we introduced a simple configuration to address this
question: it consists of an impeller rotating at a rate $\omega$
in a fluid under global rotation around the same axis at a rate
$\Omega$. This configuration was first characterized at moderate
to large Rossby number, $Ro=\omega/\Omega > 2$, in Campagne {\it
et al.}~\cite{Campagne2016}, showing a strong
two-dimensionalization of the mean flow associated with a
reduction in turbulence intensity, and hence in turbulent drag, as
$Ro$ is decreased below $\simeq O(10)$.  The aim of the present
paper is to explore the route towards two-dimensionality in this
flow at much smaller Rossby number, down to $Ro \simeq 10^{-2}$.

A closely related configuration to investigate the transition
towards two-dimensionality at low Rossby number, considered first
in the pioneering work of Hide and Titman \cite{hide_titman},
consists in spinning a disk in a rotating cylindrical container.
At sufficiently low Rossby number, apart from boundaries, the flow
is divided in two domains in solid-body rotation at angular
velocity $\Omega+\omega$ (inside the cylinder tangential to the
disk) and $\Omega$ (outside this cylinder), separated by a
Stewartson shear layer~\cite{Stewartson1957}. As the Rossby number
is further decreased, this shear layer becomes unstable with
respect to azimuthal modulations of increasing mode order
$m$~\cite{Busse1968}. This instability, often referred to as
barotropic~\cite{Vallis}, is generically observed in flows forced
by a purely axisymmetric small differential rotation in a rapidly
rotating system, in
cylindrical~\cite{niino1984,Dolzhanskii1990,konijnenberg,fruh_read}
or spherical~\cite{Hollerbach2003,Schaeffer2005} geometries.

The situation is more complex for a non-axisymmetric forcing, as
the impeller considered here. First, the natural mode of the
barotropic instability may be modified by the symmetry of the
forcing (typically the number of blades for an impeller). Second,
the non-axisymmetric ``topography'' of the forcing may be sources
of inertial waves, which may delay the transition towards pure
geostrophy. When excited by a disturbance moving at constant
velocity, these inertial waves interfere and may form a stationary
wake traveling with the disturbance~\cite{Lighthill1967,peat1976}.
In our experiment we find that, in the case of a rotating
impeller, the wakes of inertial waves attached to each blade
interact and produce small-scale fluctuations superimposed to the
large-scale geostrophic flow, making the 2D transition in such
flow richer than for a purely axisymmetric forcing.

The paper is organized as follows. Section \ref{sec:exp} describes
the experimental setup and particle image velocimetry
measurements. The large-scale flow and the azimuthal instability
as the Rossby number is decreased are described in
Sec.~\ref{sec:flow}. The dynamics of the small-scale fluctuations
are analyzed in Sec.~\ref{sec:wake}. We show in particular that
they can be described in terms of wakes of inertial waves
generated by the velocity difference between the blades and the
large-scale geostrophic flow.

\section{Experimental setup}\label{sec:exp}

\begin{figure}
    \centerline{\includegraphics[width=0.6\textwidth]{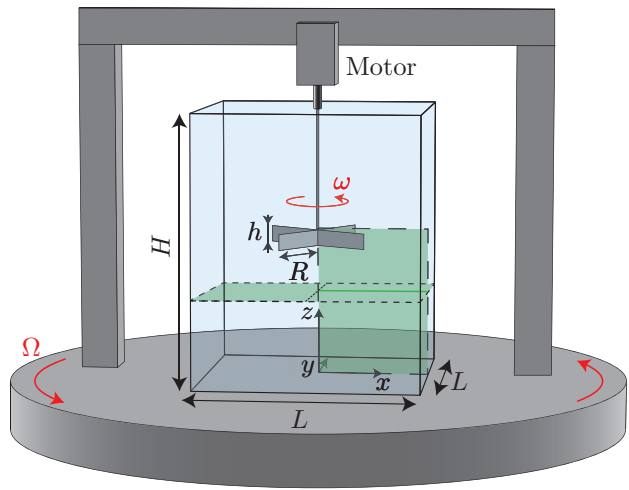}}
    \caption{Experimental setup: a motor drives an impeller at a
    constant rotation rate $\omega$ in a water-filled tank mounted on
    a platform rotating at a rate $\Omega$ (in the laboratory frame,
    the impeller spins at a rate $\omega+ \Omega$). PIV measurements
    are performed in the rotating frame, in a vertical or horizontal plane shown by the green areas.
        $H=55$ cm; $L=45$ cm; $h=3.2$ cm; $R=12$
    cm.}\label{fig:setup}
\end{figure}

The experimental setup, sketched in Fig.~\ref{fig:setup}, is
similar to the one described in Campagne \textit{et al.}
\cite{Campagne2016}, and is only briefly discussed here. A
four-parallelepipedic-blade impeller rotates at a rate $\omega =
0.5-90$~rpm around the vertical axis in a water-filled tank of $45
\times 45$~cm$^2$ square base and 55~cm height. The impeller
radius is $R=12$~cm, and each blade has a height $h=3.2$~cm and a
thickness $e=0.5$ cm. The whole system is mounted on a platform
rotating at a rate $\Omega$ in the range $4-30$ rpm. We restrict
ourselves to the regime of cyclonic rotation, i.e. when the
impeller rotates in the same direction as the platform. The Rossby
number, defined as $Ro = \omega/\Omega$, covers the range $0.017 -
3.5$, much lower than the range $2-500$ covered in
Ref.~\cite{Campagne2016}. The Reynolds number $Re=\omega R^2/\nu$
(with $\nu$ the kinematic viscosity) is in the range $10^3 -
10^5$.

We perform velocity measurements using a 2D particle image
velocimetry (PIV) system mounted on the rotating platform, either
in a vertical or a horizontal plane (green areas in
Fig.~\ref{fig:setup}). In the vertical configuration, the laser
sheet contains the impeller axis, and the imaged plane covers a
little more than one quarter of the tank section. In this
configuration, 4\,000 pairs of images of particles, separated by a
time lag chosen between $1$ and $29$~ms within a pair, are
acquired with a $2\,360 \times 1\,776$ pixels camera at 10~Hz.
Cross-correlation within pairs of images produces velocity fields
sampled on a grid of $146 \times 111$ points, with a spatial
resolution of 2.1 mm. In the horizontal configuration, the laser
sheet is located 10 cm below the impeller and almost the whole
tank section is imaged. Between 1\,500 and 15\,000 images of
particles are acquired at a rate from 3 to 30~Hz.
Cross-correlation between successive images yields velocity fields
sampled on $120 \times 102$ points with a resolution of 3.6 mm.

\section{Large-scale geostrophic flow}
\label{sec:flow}

\begin{figure}
    \centerline{\includegraphics[width=0.7\textwidth]{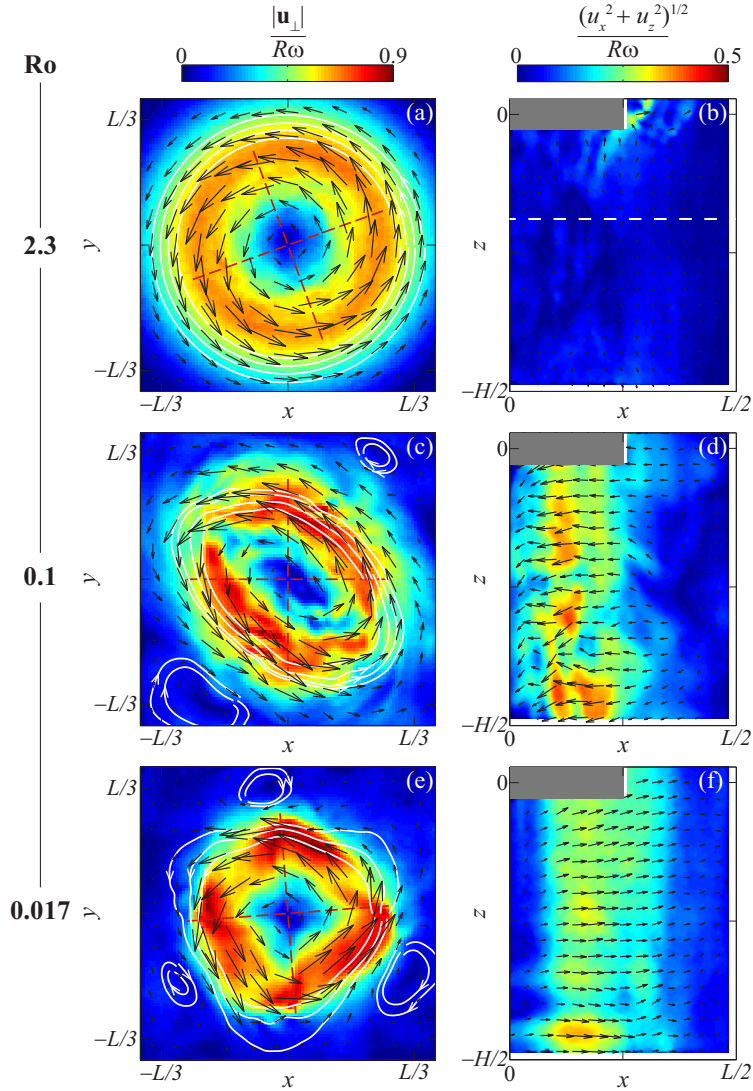}}
    \caption{Flow structure for decreasing Rossby number ($Ro=2.3$,
    $0.1$ and $0.017$). Left column: instantaneous horizontal velocity
    field and streamlines in the horizontal plane, 10~cm below the
    impeller; colormap shows the norm
    $|\mathbf{u_\perp}|=\sqrt{u_x^2+u_y^2}$; the red dashed lines show
    the position of the blades located 10~cm above. Right column:
    instantaneous velocity field in the vertical plane $y=0$; colormap
    shows the norm $\sqrt{u_x^2+u_z^2}$. The gray region shows the
    impeller blades. The dashed line in (b) shows the height of the
    plane where the horizontal measurements are performed.}
    \label{fig:velocity}
\end{figure}

The main features of the flow as the Rossby number is varied are
shown in Fig.~\ref{fig:velocity}. The horizontal structure of the
flow, visible in the velocity fields and streamlines in the left
panels, corresponds to a large circular vortex for $Ro=2.3$
[Fig.~\ref{fig:velocity}(a)]. This large-scale rotation is
subjected to a barotropic instability as the Rossby number is
decreased, leading to an azimuthal modulation of increasing order
$m$: The intermediate value $Ro=0.1$ [Fig.~\ref{fig:velocity}(c)]
shows an elliptical vortex (azimuthal mode $m=2$), surrounded by
two counter-rotating satellite vortices, whereas the lowest value
$Ro=0.017$ [Fig.~\ref{fig:velocity}(e)] shows a triangular vortex
(azimuthal mode $m=3$) surrounded by three satellite vortices.
Such azimuthal modulation with increasing mode $m$ as $Ro$ is
decreased is a classical feature of barotropic instabilities in
geostrophic flows~\cite{Vallis}, also found with a purely
axisymmetric forcing such as in the flat-disk configuration of
Hide and Titman~\cite{hide_titman,Busse1968}. For the low Rossby
numbers considered here, the flow is approximately vertically
invariant, as shown in the vertical cuts at $y=0$ in the right
panels of Fig.~\ref{fig:velocity} (this vertical invariance cannot
be checked however in the particular case of the purely
axisymmetric flow, for $Ro=2.3$, because only the velocity
components in the vertical plane are measured here). In
particular, no poloidal recirculation can be seen, in contrast to
the large Rossby number case in which a radial ejection in the
impeller plane and a pumping along the vertical axis take place
(see, e.g., Fig.~4 in Ref.~\cite{Campagne2015}). Here such
poloidal recirculation, which is incompatible with the
Taylor-Proudman theorem in the limit of vanishing Rossby number,
is already strongly inhibited at $Ro \simeq O(1)$.

\begin{figure}
    \centerline{\includegraphics[width=0.7\textwidth]{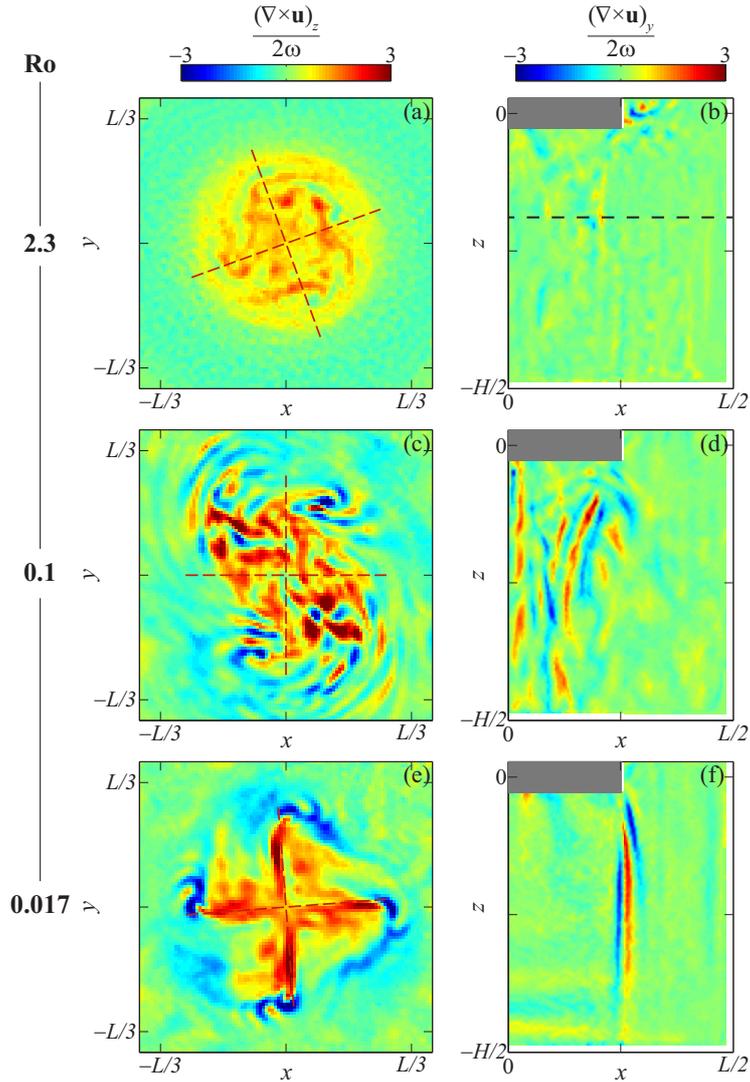}}
    \caption{Vorticity fields for decreasing Rossby number
    ($Ro=2.3$, $0.1$ and $0.017$), revealing the small-scale features
    of the flow. Left column: instantaneous vertical vorticity
    component $(\mathbf{\nabla} \times \mathbf{u})_z$ at the same
    times than in the left column of Fig.~\ref{fig:velocity}. Right
    column: out-of-plane vorticity component $(\mathbf{\nabla} \times
    \mathbf{u})_y$ in the vertical plane $y=0$ [not at the same time
    than in the right column of Fig.~\ref{fig:velocity} for (d) and
    (f)].}\label{fig:vorticity}
\end{figure}

In addition to the large-scale geostrophic flow, we also observe
small-scale three-dimensional fluctuations, mostly visible in the
vorticity fields shown in Fig.~\ref{fig:vorticity}. These
vorticity fluctuations are stronger at the tip of the impeller
blades for $Ro=2.3$ [Fig.~\ref{fig:vorticity}(b)], but they are
also visible in the region below the impeller
[Fig.~\ref{fig:vorticity}(a)]. As the Rossby number is decreased,
vorticity sheets gradually appear below the impeller. While these
vorticity sheets are disordered and fill the whole region below
the impeller for $Ro=0.1$ [Figs.~\ref{fig:vorticity}(c-d)], they
become almost vertically-invariant and closely follow the shape of
the impeller for $Ro=0.017$. This vertical invariance, expected in
the limit of small Rossby number, is illustrated by the red cross
reproducing the shape of the impeller 10~cm below it in
Fig.~\ref{fig:vorticity}(e). We further describe the dynamics of
these vorticity sheets in Sec.~\ref{sec:wake}.

\begin{figure}
\centerline{\includegraphics[width=0.95\textwidth]{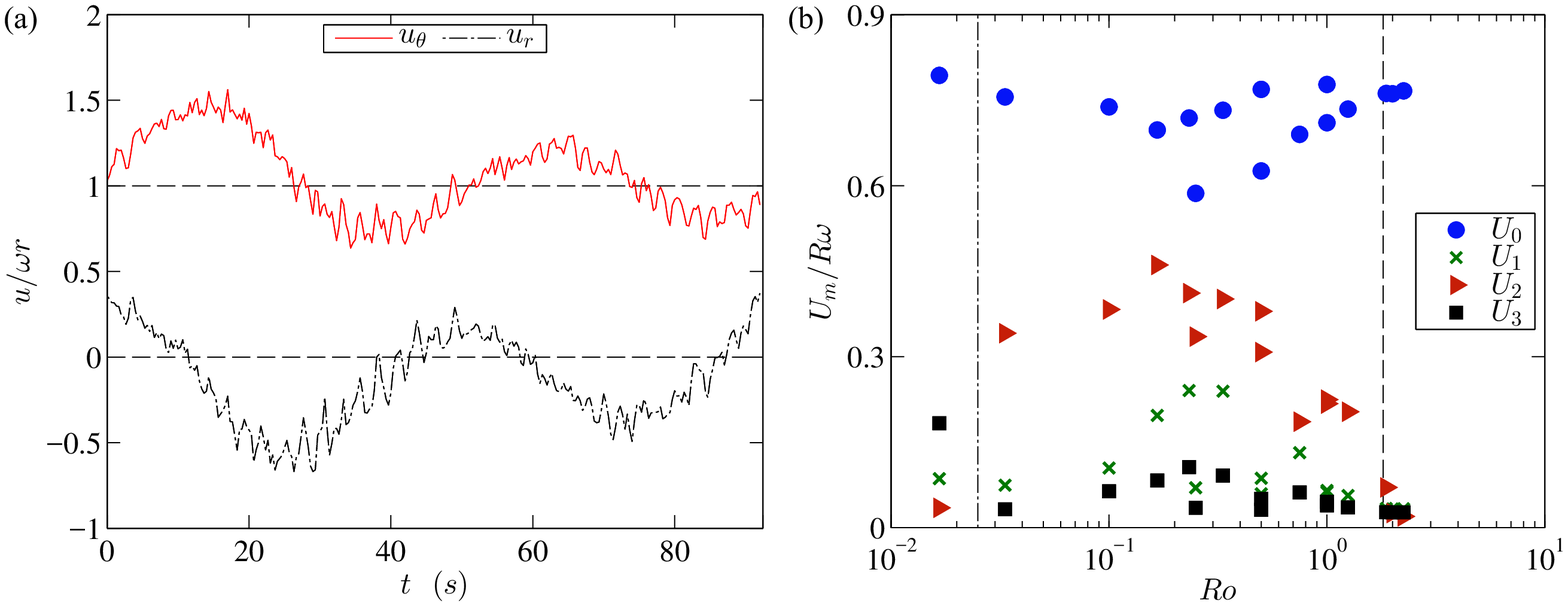}}
\caption{(a) Phase-averaged azimuthal and radial velocities,
$u_\theta$ and $u_r$, as a function of time for $Ro=0.1$ at point
$(x,y)= (0.4R,0)$. The main oscillation corresponds to the slow
rotation of the elliptical mode $m=2$ at the angular phase
velocity $\omega_2$. The rapid oscillations correspond to the
blade crossing frequency $4\omega$. (b) Modal amplitudes $U_m$ of
the most energetic modes $m=0-3$ as a function of the Rossby
number. Vertical lines show the transition Rossby numbers
$Ro_c^{(2-3)}$ between modes $m=2$ and 3 and $Ro_c^{(0-2)}$
between modes $m=0$ and 2.} \label{fig:Um}
\end{figure}

For low Rossby number, the pattern of the large-scale
non-axisymmetric geostrophic flow slowly rotates as a whole in the
same direction as the impeller with a well-defined angular phase
velocity. This is illustrated in Fig.~\ref{fig:Um}(a), showing the
time evolution of the azimuthal and radial velocity components at
point $(x,y)=(0.4R,0)$ during one revolution of the elliptical
vortex for $Ro=0.1$. The azimuthal component $u_\theta / \omega r$
oscillates around 1 and the radial component $u_r / \omega r$
oscillates around 0, both with an amplitude of order of 0.5. This
indicates that, even when the axisymmetry is broken, the
azimuthally averaged flow still essentially rotates at the angular
velocity imposed by the impeller. The slow modulation shown in
Fig.~\ref{fig:Um}(a) is dominated by a mode $m=2$, but it also
contains a weak mode $m=1$ (the first maximum is more pronounced
than the second one), corresponding to a circular translation of
the vortex axis.

We have systematically computed the modal contributions of the
geostrophic flow as a function of the Rossby number. For this, we
first compute for each radius $r$ and time $t$ the energy
${\cal U}^2_m(r,t)$ of the different modes $m$ from the
power spectrum of the angular profile of the horizontal velocity
${\bf u}_\perp(r, \theta, t)$. The modal amplitude is finally obtained
from a spatial (in cylindrical coordinate) and temporal average of
${\cal U}^2_m(r,t)$,
\begin{equation}
U_m = \left< \frac{4}{R^2}  \int_0^R {\cal U}^2_m (r,t) r dr
\right>^{1/2}, \label{eq:Um}
\end{equation}
where $\langle ~ \rangle$ is the time average. Because of the
non-axisymmetric boundaries of the square tank, the radial
integration cannot be extended over the whole domain and is
arbitrarily truncated at $r=R$. The normalization in
Eq.~(\ref{eq:Um}) is chosen such that a solid-body rotation of
angular velocity $\omega$ extending up to $r=R$ yields $U_0=\omega
R$ and $U_{m}=0$ for $m>0$.

The amplitude $U_m$ of the most energetic modes $m=0-3$ are
plotted in Fig.~\ref{fig:Um}(b) as a function of the Rossby
number. The axisymmetric contribution $m=0$ is always dominant,
with a nearly constant amplitude, $U_0 / \omega R \simeq 0.6-0.8$:
these values reflect the fact that the solid-body rotation
component of the flow only extends up to $r\simeq0.7R$ (see Fig.~5
in Ref.~\cite{Campagne2016}). The amplitude of the mode $m=1$
(circular translation of the vortex core around the symmetry axis)
remains moderate at all Rossby numbers, $U_1/\omega R \simeq
0.1-0.2$. Among the non-axisymmetric modes, the elliptical mode
$m=2$ is dominant over a wide range, $Ro_c^{(2-3)} < Ro <
Ro_c^{(0-2)}$, with transition Rossby numbers estimated as
$Ro_c^{(2-3)} = 0.025 \pm 0.01$ and $Ro_c^{(0-2)} = 1.75\pm 0.25$,
while the triangular mode $m=3$ becomes dominant only for the
smallest explored Rossby number $Ro=0.017$. Note that the data for
different background rotations $\Omega$ in Fig.~\ref{fig:Um}(b)
suggest that the  transition Rossby numbers $Ro_c^{(2-3)}$ and
$Ro_c^{(0-2)}$ do not significantly depend on $\Omega$. Such
critical Rossby numbers are however not expected to remain
constant with $\Omega$: a weak dependence with respect to the
Ekman number, $Ek= \nu / \Omega R^2$, is usually found in
rotating-disk experiments~\cite {hide_titman,niino1984,fruh_read}.

\begin{figure}
    \centerline{\includegraphics[width=.6\textwidth]{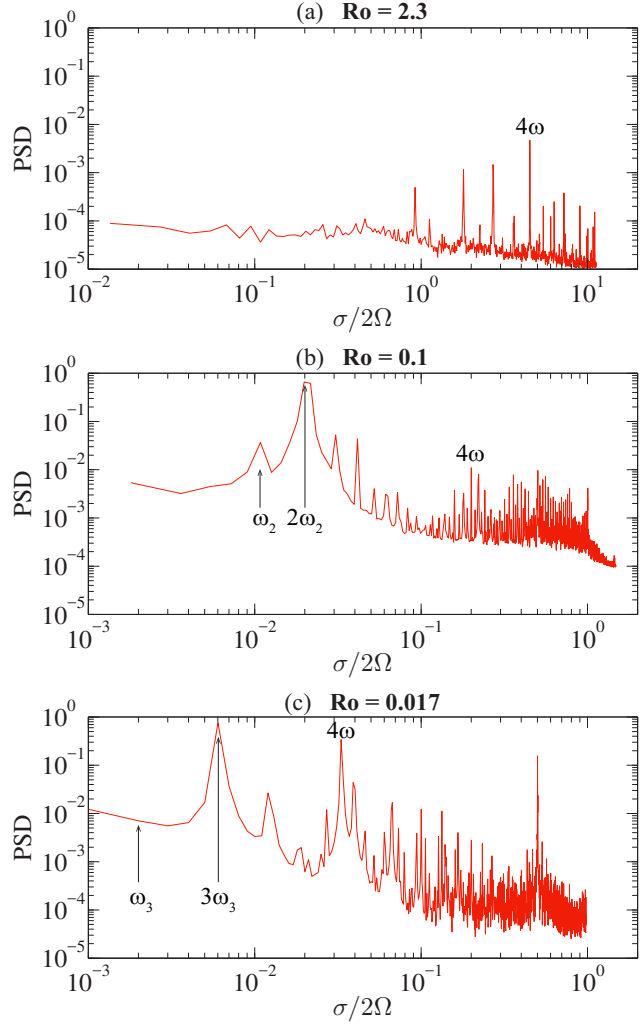}}
    \caption{Temporal power spectra of the horizontal velocity field
    for the three Rossby numbers of Fig.~\ref{fig:velocity} ($Ro=2.3$,
    $Ro=0.1$ and $Ro=0.017$), showing the growth of low-frequency
    peaks associated to the angular drift of the barotropic modes
    $m=2$ and $m=3$. The frequency $\sigma$ is normalized by the
    background vorticity $2\Omega$.} \label{fig:spectra}
\end{figure}

The angular phase velocity $\omega_m$ is further determined for
each mode from a temporal Fourier analysis of the velocity
measured in the horizontal plane. For all Rossby numbers, the
power spectra, reported in Fig.~\ref{fig:spectra}, present a peak
at frequency $4\omega$ corresponding to the blades motion. In
addition, for $Ro<Ro_c^{(0-2)}$ [Figs.~\ref{fig:spectra}(b,c)],
low-frequency peaks of larger amplitude appear: they correspond to
the rotation of the non-axisymmetric geostrophic pattern. The
fundamental frequency $\omega_m$ is only visible for $m=2$ because
of the too small sampling duration for $m=3$, but its harmonic $m
\omega_m$, which contains most of the energy, is well defined both
for $m=2$ and $m=3$ [Figs.~\ref{fig:spectra}(b) and (c)]. The
coupling of the rapid oscillations at the blade frequency
$4\omega$ with the slow barotropic modulation at frequency $m
\omega_m$ also produces a series of high-frequency harmonics at $4
\omega \pm N m \omega_m$, with integer $N$, surrounding the blade
frequency.

\begin{figure}
    \centerline{\includegraphics[width=.95\textwidth]{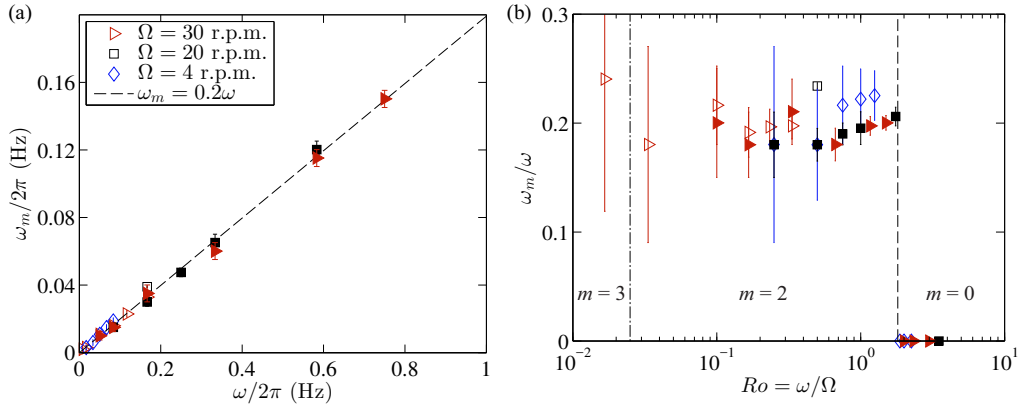}}
    \caption{(a) Angular phase velocity $\omega_m$ of the azimuthal
    modes $m=2$ and 3 as a function of the impeller rotation rate $\omega$, for
    several platform rotation rates $\Omega$; the dashed line shows a
    linear fit $\omega_m=0.2\omega$. (b) Ratio $\omega_m/\omega$
    versus Rossby number ($\omega_m$ is set to 0 for the axisymmetric
    mode $m=0$). Vertical lines show the transition Rossby numbers
    $Ro_c^{(2-3)}$ and $Ro_c^{(0-2)}$. Open (resp. filled) symbols
    correspond to frequency measurements in the horizontal (resp.
    vertical) plane. Error bars correspond to the frequency resolution
    of the power spectral density.}
        \label{fig:wm}
\end{figure}

The angular frequency $m \omega_m$ is measured for $m=2$ and $3$
from the maxima in the power spectra, and the corresponding
angular phase velocity $\omega_m$ is reported in Fig.~\ref{fig:wm}
as a function of $\omega$. We observe a linear relation, $\omega_m
\simeq 0.2 \omega$, for both $m=2$ and $3$ (note that $m=3$ is
present only for the lowest Rossby number, $Ro=0.017$). This
constant frequency ratio $\omega_m / \omega$, also observed in
rotating-disk experiments \cite{hide_titman,fruh_read}, indicates
that the vortex pattern, even if composed as a superposition of
modes, essentially rotates as a whole. The ratio $\omega_m /
\omega \simeq 0.2$ neither varies significantly with the
background rotation $\Omega$ [see Fig.~\ref{fig:wm}(b); in this
plot we set $\omega_m=0$ for the axisymmetric mode $m=0$], here
again in qualitative agreement with rotating-disk experiments.

\section{Wakes of Inertial waves}\label{sec:wake}

\begin{figure}
    \centerline{\includegraphics[width=0.95\textwidth]{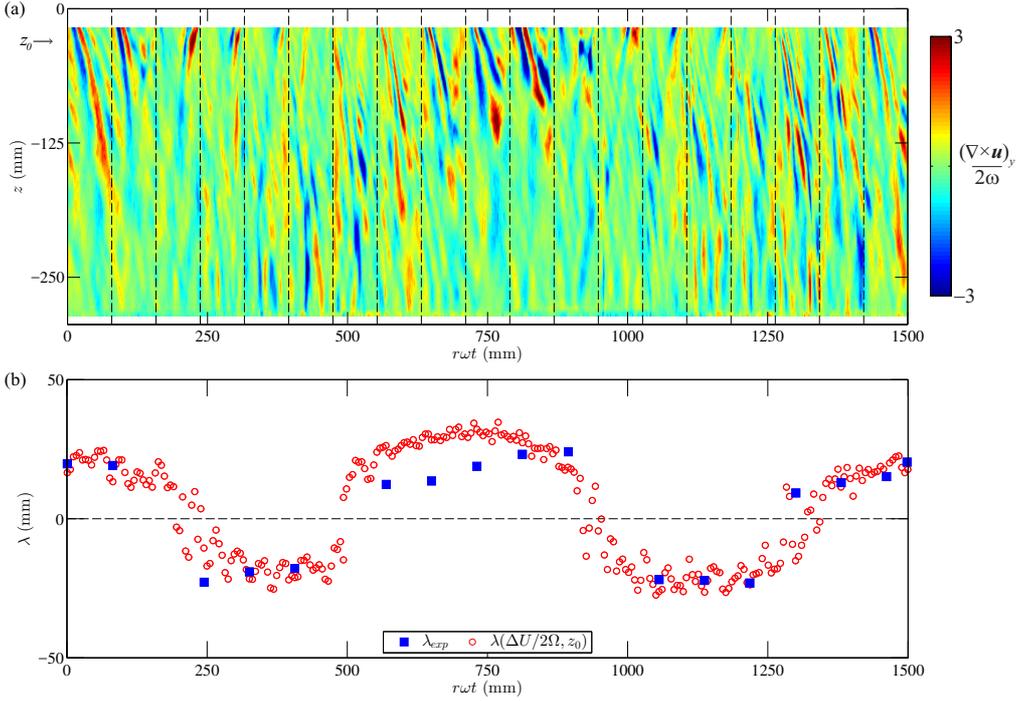}}
    \caption{(a) Spatio-temporal diagram along $z$ of the azimuthal
    component of the vorticity $(\nabla \times \mathbf{u})_y$ at
    $Ro=0.1$, $r=x=0.4R$. The dashed lines indicate the crossings of the
        blades through the measurement plane. Time is
    expressed in terms of an effective azimuthal coordinate $r\omega t$. (b)
    Blue squares: horizontal wavelength in the vorticity packets induced by each
    blade crossing the measurement plane, measured at height
    $z_0=-30$ mm. This wavelength is computed from the vorticity autocorrelation
        over a window along the coordinate $r \omega t$,
        and its sign is set to negative when the
    wake precedes the blade. Red circles: predicted wavelength from
        Eq.~(\ref{eq:isophyx}), computed using
    the instantaneous azimuthal velocity difference between the fluid
    and the blade $\Delta U_\theta$.}
    \label{fig:std}
\end{figure}

We now characterize the small-scale structures observed below (and
above) the impeller when the azimuthal modulation of the
large-scale flow is present. Figure~\ref{fig:std}(a) shows the
temporal evolution of the vertical profile of the azimuthal
vorticity $(\nabla \times \mathbf{u})_y$ at radius $r=x=0.4R$ in
the case $Ro=0.1$. In this diagram the time is expressed in terms
of an equivalent azimuthal coordinate $r\omega t$. We can see that
almost every time a blade crosses the measurement plane (shown by
vertical dashed lines) a packet of vorticity sheets of alternate
sign is emitted. The thickness of the vorticity sheets along the
azimuthal coordinate $r\omega t$, of order of 10 mm, is consistent
with their thickness in the radial direction visible in
Fig.~\ref{fig:vorticity}(d). These vorticity packets are
alternatively following and preceding the crossings of the blades,
on a slow time scale given by the angular phase velocity of the
geostrophic flow $m\omega_m$ (with $m=2$ and $\omega_2 \simeq 0.2
\omega$).

We show in the following that these packets of vorticity sheets
correspond to wakes of inertial waves emitted by the blades. Such
wakes originate from the azimuthal modulation of the geostrophic
flow, generating a slowly evolving velocity difference
\begin{equation}
\Delta {\bf U}(r,\theta,t)=\omega r {\bf e}_\theta -{\bf u}(r,\theta,t)
\label{eq:du}
\end{equation}
between the blades and the fluid below and above the impeller. The
radial and azimuthal components of $\Delta {\bf U}$ oscillate
around 0 at frequency $m\omega_m$ [see Fig.~\ref{fig:Um}(a)]. For
instance, in the case $m=2$, when $\Delta U_\theta>0$, the blade
is along the major axis of the ellipse and therefore rotates
faster than the fluid, generating a wake behind it. On the
contrary, when $\Delta U_\theta<0$, the blade is along the minor
axis and rotates slower than the fluid, so the wake precedes the
blade. In the intermediate cases, the velocity difference $\Delta
{\bf U}$ is essentially radial and no clearly defined wake can be
observed.

\begin{figure}
    \centerline{\includegraphics[width=.6\textwidth]{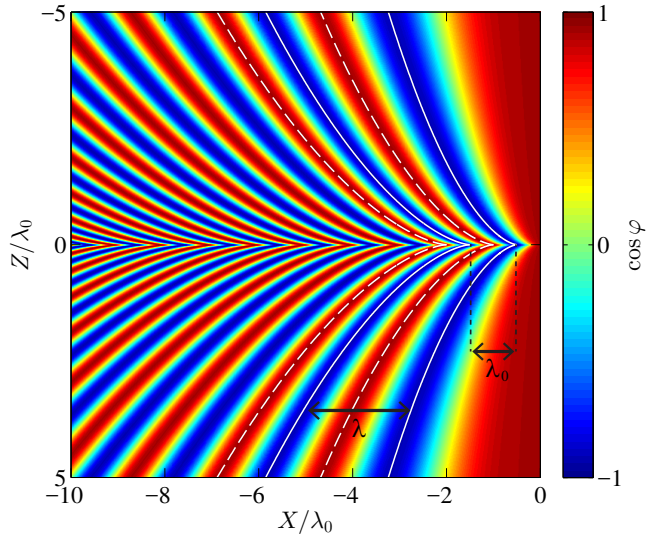}}
    \caption{Phase field $\cos \varphi$ (\ref{eq:isophyx}) of
        the linear inviscid
    wake of inertial waves produced by the translation of a line
    source in the direction $X>0$ at a constant velocity $U$. The
    axis are normalized by the natural
    wavelength $\lambda_0$ (\ref{eq:l0}) along the wake axis.
    White lines show lines of constant phase $\varphi=\pi$,
    $2\pi$, $3\pi$, and $4\pi$ from right to left.
    } \label{fig:phase}
\end{figure}

In order to confirm this picture, we can refer to the simple model of
a linear inviscid wake of inertial waves produced by the
translation of a source, analyzed by Lighthill~\cite{Lighthill1967} and
Peat \& Stevenson~\cite{peat1976}. We consider a line source, invariant along $Y$,
traveling at constant velocity $U$ in the direction $X>0$ normal to the
rotation axis $Z$. In the frame of the source, the phase field $\varphi(X,Z)$
of the wake is such that the lines of constant phase $\varphi$ write
\begin{eqnarray}
\begin{cases}
X=-\lambda_0 \dfrac{\varphi}{2\pi}\dfrac{2-\cos^2\theta}{\cos\theta},\label{eq:isophyx}\\
Z=\lambda_0 \dfrac{\varphi}{2\pi}\dfrac{\sin^3\theta}{\cos^2\theta},
\end{cases}
\end{eqnarray}
where
\begin{equation}
\lambda_0=\pi  U/\Omega
\label{eq:l0}
\end{equation}
is the natural wavelength along the source trajectory,
and $\theta$ a parameter ranging from $0$ to $\pi/2$. The wave field
$\cos\varphi$ is shown in Fig.~\ref{fig:phase} in normalized
coordinates $(X,Z)/\lambda_0$. The wavelength $\lambda_0$ is $2\pi$ times
the gyration radius $U/2\Omega$ of a fluid particle oscillating horizontally
at velocity $U$ and natural frequency $2\Omega$ in the inertial wave.

In order to compare the shape of the wave packets observed in
Fig.~\ref{fig:std}(a) to the phase lines of this simplified model,
we approximate the blade motion locally as a linear translation
(we therefore neglect the curvature of the trajectory), so that we
can identify in Eqs.~(\ref{eq:isophyx})-(\ref{eq:l0}) the source
velocity $U$ to the (slowly varying in time) azimuthal component
$\Delta U_\theta$ of the velocity difference between the blade and
the fluid, and the coordinate $X$ to the equivalent azimuthal
coordinate $r \omega t$.

For each wave beam, we first compute the characteristic wavelength
$\lambda_{exp}$ at a fixed distance $z_0=-30$~mm below the
impeller. We define $\lambda_{exp}$ as twice the scale
corresponding to the first minimum of the vorticity
autocorrelation function computed over a window of 80~mm along
$X=r\omega t$ and centered on the wave beam of interest. We set
the sign of $\lambda_{exp}$ as positive when the wake follows the
blade and negative when it precedes it. This characteristic
wavelength is plotted in Fig.~\ref{fig:std}(b), and compared with
the predicted wavelength from Eq.~(\ref{eq:isophyx}), evaluated
for the same value of $Z$ and for the azimuthal component $\Delta
U_\theta$ of the velocity difference $\Delta {\bf U}$ at the same
time. In spite of the simplicity of the model, we obtain a
reasonable agreement between the measured and the predicted
wavelength. The discrepancies can be ascribed to the various
assumptions of the model which are not satisfied in the
experiment: in addition to the curvature of the source trajectory,
the model neglects the finite size of the source, the radial
component of the velocity difference $\Delta {\bf U}$, and the
variation with time of $\Delta U_\theta$.

\begin{figure}
    \centerline{\includegraphics[width=0.95\textwidth]{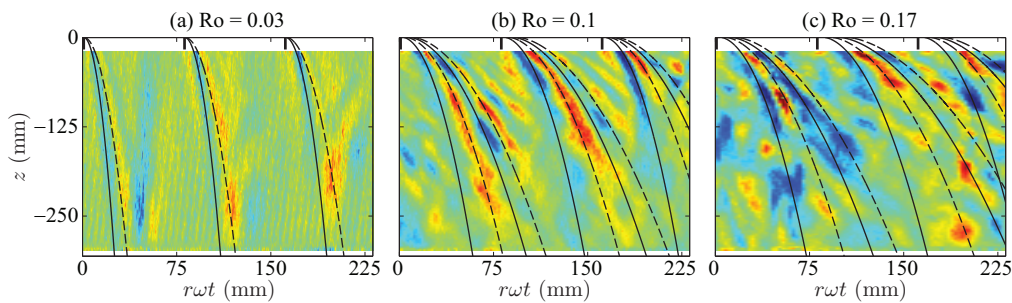}}
\caption{Spatio-temporal diagrams of the azimuthal component of
the vorticity along $z$ at $(x,y) = (0.4R,0)$ for three Rossby
numbers. Black lines show lines of constant phase $\varphi=\pi$,
$2\pi$, $3\pi$, and $4\pi$ (from left to right) given by
Eq.~(\ref{eq:isophyx}). For each wake, the origin $(X=0,Z=0)$ of
the reference frame of the phase lines in Eq.~(\ref{eq:isophyx})
is taken at the center of the blade, at the time given by the
blade crossing the measurement plane. The value of $\lambda_0$ in
Eq.~(\ref{eq:isophyx}) is computed from the instantaneous
azimuthal velocity difference $\Delta U_\theta$ between the fluid
and the blade.} \label{fig:stdphase}
\end{figure}

We further test our description by superposing the predicted phase
lines (\ref{eq:isophyx}) on the spatio-temporal diagrams in
Fig.~\ref{fig:stdphase} for three values of the Rossby number,
$Ro=0.03$, $0.1$ and $0.17$, for which the dominant mode is $m=2$.
Three blade crossings are shown for each $Ro$, chosen during time
intervals for which $\Delta U_\theta$ is positive (i.e., when the
wakes follow the blades) and approximately constant. Here again a
reasonable match is obtained, at least for $Ro=0.03$ and 0.1,
which confirms that the small-scale fluctuations at low $Ro$
correspond to wakes of inertial waves emitted by the blades. As
expected the vorticity field becomes increasingly disordered as
$Ro$ increases, as the result of instabilities in the wakes or
interactions of the wakes with the turbulent structures directly
produced by the blade motion, and the match with the model
(\ref{eq:isophyx}) becomes worse as $Ro$ approaches $O(1)$.

\section{Conclusion}\label{sec:concl}

We have characterized in this paper the two-dimensionalization
process in the flow produced by a slowly rotating impeller in a
rapidly rotating fluid, extending the range of Rossby numbers of
Campagne {\it et al.}~\cite{Campagne2016} down to $10^{-2}$.
Although two-dimensionality is already satisfied for the mean flow
at $Ro \simeq O(1)$, the small scales remain three-dimensional and
disordered down to $10^{-2}$. In this regime the flow  can be
described as the superimposition of a large-scale azimuthally
modulated geostrophic flow and small-scale vorticity sheets, which
correspond to wakes of inertial waves originating from the
velocity difference between the impeller and the non-axisymmetric
geostrophic flow. This situation is expected in general for flows
produced by a non-axisymmetric rotating device, in contrast to
experiments with axisymmetric forcing, as in Hide \&
Titman~\cite{hide_titman}.

We expect that this regime of disordered wakes at Rossby number
$Ro \simeq 10^{-2}-1$ is a transient in the route towards pure
two-dimensionalization: as the Rossby number is further decreased,
the barotropic instability should lead to azimuthal modulations of
increasing order $m$, associated with a decreasing velocity
difference between the geostrophic flow and the impeller. In this
limit the wakes produced by the blades should become vertical, a
tendency already visible in Fig.~\ref{fig:vorticity}(f),
protruding the shape of the impeller to $z \rightarrow \pm
\infty$, eventually leading to a pure two-dimensional flow at
vanishing Rossby number. Interestingly, the disordered wake regime
observed here may form, at larger Reynolds number, a particular
state of inertial wave
turbulence~\cite{Galtier2003,Cambon2004,Nazarenko}: the localized
wave generation implies an upward and downward energy propagation
on each side of the impeller, leading to a separation of the sign
of helicity~\cite{Ranjan2014}.

\section*{Acknowledgements}

We acknowledge A. Campagne and B. Gallet for fruitful discussions, and
J. Amarni, A. Aubertin, L. Auffray and R. Pidoux
for experimental help. This research was funded by Investissements
d'Avenir LabEx PALM (ANR-10-LABX-0039-PALM). FM acknowledges the
Institut Universitaire de France.

\end{document}